# Twisted Light Transmission over 143 kilometers

Mario Krenn[1,2,*], Johannes Handsteiner[1,2], Matthias Fink[2], Robert Fickler[1,2,3], Rupert Ursin[2], Mehul Malik[1,2], Anton Zeilinger[1,2,*]

[1]Vienna Center for Quantum Science and Technology (VCQ), Faculty of Physics,
University of Vienna, Boltzmanngasse 5, A-1090 Vienna, Austria.
[2]Institute for Quantum Optics and Quantum Information (IQOQI),
Austrian Academy of Sciences (ÖAW), Boltzmanngasse 3, A-1090 Vienna, Austria.
[3]Department of Physics and Max Planck Centre for Extreme and Quantum Photonics, University of Ottawa, Ottawa, K1N 6N5, Canada.
*correspondence to mario.krenn@univie.ac.at and anton.zeilinger@univie.ac.at

**Spatial modes of light can potentially carry a vast amount of information, making them promising candidates for both classical and quantum communication. However, the distribution of such modes over large distances remains difficult. Intermodal coupling complicates their use with common fibers, while free-space transmission is thought to be strongly influenced by atmospheric turbulence. Here we show the transmission of orbital angular momentum modes of light over a distance of 143 kilometers between two Canary Islands, which is 50 times greater than the maximum distance achieved previously. As a demonstration of the transmission quality, we use superpositions of these modes to encode a short message. At the receiver, an artificial neural network is used for distinguishing between the different twisted light superpositions. The algorithm is able to identify different mode superpositions with an accuracy of more than 80% up to the third mode order, and decode the transmitted message with an error rate of 8.33%. Using our data, we estimate that the distribution of orbital angular momentum entanglement over more than 100 kilometers of free space is feasible. Moreover, the quality of our free-space link can be further improved by the use of state-of-the-art adaptive optics systems.**

The transverse spatial modes of light offer an additional degree of freedom for encoding information in both classical and quantum communication. In classical communication, such modes can be used for multiplexing information and for increasing the achievable data rate per frequency and polarization channel [1-4]. In quantum information science they are a physical realization of a high-dimensional quantum state [5-9]. Such states allow one to encode more than one bit of information per photon. The large state space can increase the channel capacity and improve robustness to eavesdropping and noise [10, 11] in quantum communication schemes [12-15]. The distribution of photons carrying different spatial modes over macroscopic distances is thus essential for both quantum and classical applications, as well as for fundamental tests of quantum mechanics. While significant progress has been made in fiber-based solutions [16, 17], these methods are still in their infancy. Here we focus on a different way distributing such modes, namely long-distance transmission through free space. This is relevant in situations where fibers are not applicable, such as for long-distance quantum communication and communication with satellites.

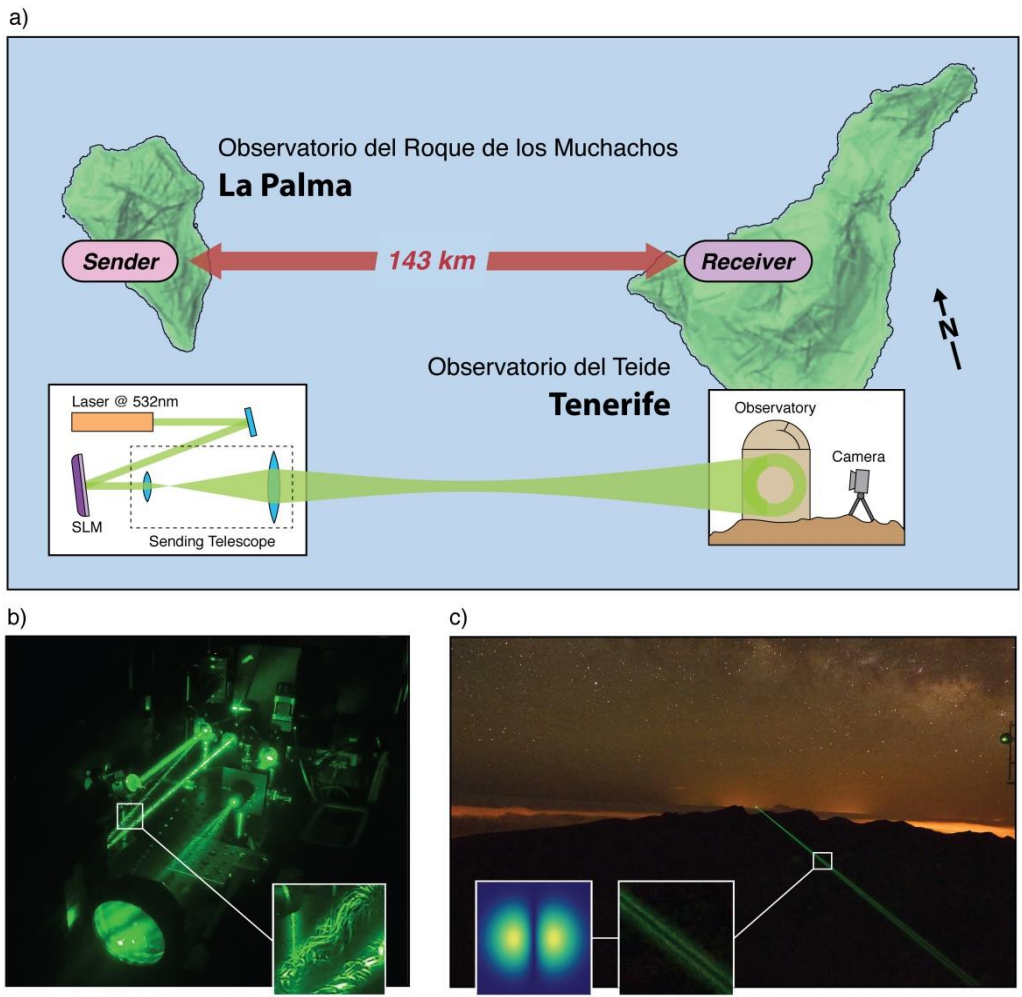

**Figure 1: a)** Sketch of the experimental setup. The sender is located on the roof of the Jacobus Kapteyn telescope on the island of La Palma and consists of a 60mW laser with a wavelength of 532nm modulated by a spatial light modulator (SLM). Different phase holograms on the SLM encode different spatial modes. The modes are magnified with a sending telescope and sent over 143 km to the receiver at the island of Tenerife. Extra mirrors used in the actual sending setup are not shown for simplicity. The structure of received modes is observed on the wall of the telescope "Optical Ground Station" (OGS) owned by the European Space Agency and recorded with a camera. **b)** A photo of the sender taken during extremely turbulent conditions on La Palma. Small vortices and eddies formed by the water vapor in the air are clearly visible in the inset. At the sending lens an $\ell = \pm 1$ can be seen. Modes sent under these conditions were not discernable at the receiver. **c)** Long-time exposure photo showing an OAM superposition of $\ell = \pm 1$ being transmitted over the Caldera de Taburiente (silhouetted in black) from La Palma to Tenerife. The insets show that the double-lobed modal structure of the beam is clearly visible. A theoretical plot of the mode superposition cross-section is shown for comparison.

Atmospheric turbulence plays a significant role in the free-space transmission of spatial modes. These effects of the atmosphere have been investigated in many recent theoretical studies [18-24] and lab-scale simulations [25-30]. While these investigations clearly show that transmitting spatial modes of light over large distances is very challenging, several experimental investigations of free-space long-distance transmission of spatial modes have been successfully carried out recently. Most of them utilize so-called twisted light modes, where the phase front of the light has a helical phase structure. As these modes can carry an integer number of orbital angular momentum (OAM) quanta, they are often referred to as OAM modes. In 2012, a classical communication experiment with twisted radio waves was performed over ~420 meters in Venice [31]. Single photons carrying OAM in the visible frequency were transmitted over ~210 meters in a QKD experiment in Padua [32]. The experiment was carried out inside a large hall as light in the visible frequency is significantly more influenced by the turbulence because of shorter wavelengths. More recently, sixteen different spatial modes were used to encode information for classical communication over an intra-city 3-kilometer link across Vienna [33]. In the same 3-kilometer link, quantum entanglement encoded in the OAM degree of freedom was transmitted using the first two higher order OAM modes [34], demonstrating that single-photon spatial coherence and two-photon coherence of spatial modes survive in a turbulent long-distance link. One experiment in Erlangen tested classical transmission and cross-talk of OAM beams over 1.6 kilometers [35]. Finally, a high-speed classical communication experiment using OAM mode multiplexing was performed over 120 meters of free-space in Los Angeles, transmitting 400 GBit/sec [36].

In this work, we test the effect of atmospheric turbulence on a 143-kilometer OAM free-space link between two Canary Islands, La Palma and Tenerife. In doing so, our experiment increases the maximum distance achieved in a free-space OAM link by a factor of 50. We use OAM mode superpositions of $\ell=\pm1$, $\pm2$ and $\pm3$ with different relative phases for encoding information. The relative phases result in a rotation of the mode structure, which allows these modes to be distinguished according to their intensity. This technique has previously been exploited to investigate the transmission quality in both a classical [33] and a quantum experiment [34] in an intra-city 3-kilometer link. For characterization of the received mode quality, we record images of the intensity distribution observed on the white wall of the telescope "*Obervatorio del Teide*" and analyze them with a pattern recognition algorithm based on an artificial neural network. By calculating the cross-talk, we find that the modal structure can be distinguished quite well even without the use of any adaptive optics correcting for the effects of atmospheric turbulence. Finally, to visualize the quality of the transmission, we use these modes to encode, transmit, and recover a short message.

In the experiment, we use Laguerre-Gauss (LG) modes of light that are characterized by a spiral phase distribution $\exp(i\ell\varphi)$, where $\ell$ stands for the orbital angular momentum or topological charge of the photons [37]. The radial mode number n is zero in all experiments. In the center of the beam, a phase singularity leads to an intensity null along the beam axis, which gives OAM modes a ring-shaped intensity pattern. Equally weighted superpositions of LG modes with opposite OAM can be written as

$$LG_{\pm\ell}^{\alpha}(r,\varphi) = \frac{1}{\sqrt{2}}\left(LG_{+\ell}(r,\varphi) + e^{i\alpha}LG_{-\ell}(r,\varphi)\right), \qquad (1)$$

where $\alpha$ denotes the relative phase between the two modes. The transverse phase is radially uniform and has $2\ell$ phase jumps of $\pi$ in the azimuthal direction, which leads to $2\ell$ maxima and minima arranged symmetrically in a ring. The phase $\alpha$ is directly related to the angular position of the structure $\gamma = \frac{360°}{2\pi}\frac{\alpha}{2l}$. Hence, a simple determination of the angular position, e.g. by recording the intensity structure, can be used to reveal the relative phase in classical [33] and even in quantum experiments [34, 38, 39].

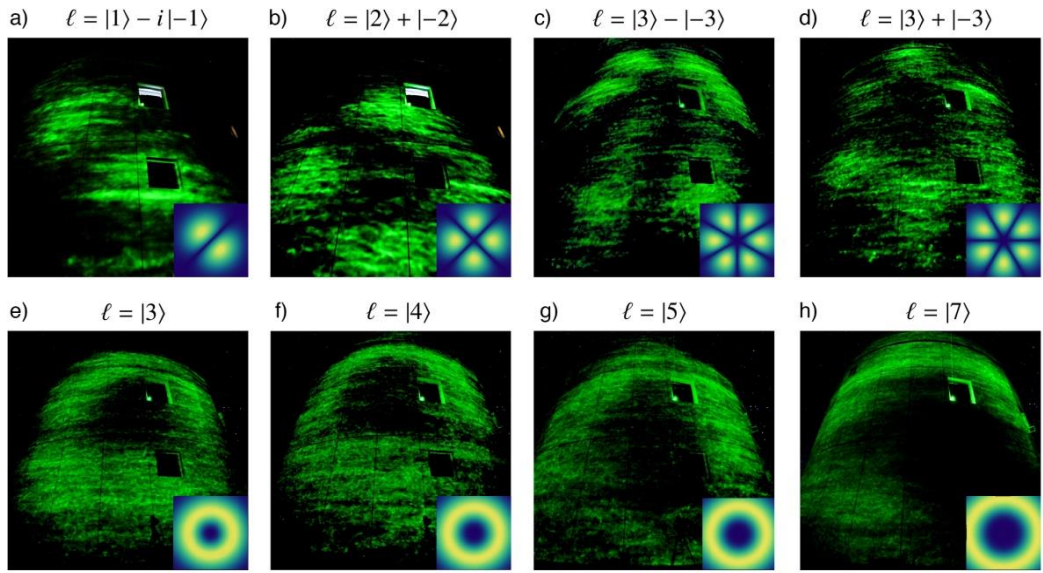

**Figure 2: a)-d)** Examples of OAM mode superpositions received on the wall of the telescope "Observatorio del Teide" after propagating through 143km of free space between the islands of La Palma and Tenerife. The lobed modal structure is clearly visible for mode superpositions with $\ell=\pm1$, $\pm2$, and $\pm3$. Images c) and d) show the rotation of a $\ell=\pm3$ mode superposition by $\pi/3$ when the relative phase $\alpha$ is changed by $\pi$. **e)-h)** Examples of pure OAM (vortex) modes observed at the receiver. The intensity null at the center of the modes is clearly visible. The mode diameter gets larger as the OAM quantum number $\ell$ is increased, and is seen to approach the size of the telescope wall for $\ell=7$. The size of the modes clearly increases for higher orders. Note that these images were taken at a time when atmospheric conditions were stable.

A schematic of the experimental setup for sending and receiving the spatial modes can be seen in Fig. 1a. At the sending station on La Palma, a green laser with a wavelength of 532nm and a power of 60mW was used for encoding the twisted light modes and their superpositions. The laser was modulated with a phase-only spatial light modulator (SLM), which imprinted the spatial modes onto the beam. Then, it was magnified with a telescope to approximately 4 cm diameter, and transmitted with a high-quality f=28 cm lens through 143km of free space to the island of Tenerife, where the receiver was located. There, the mode structure was observed on the white wall of an observatory and recorded with a NIKON D3S camera with varying exposure times. A beacon laser shining back from Tenerife to La Palma was used for initial alignment. However, no active tracking or adaptive optics was used for correcting the atmospheric turbulence during the data transmission itself. The recorded images were then analyzed to recover the encoded information in an automatized manner.

Fig. 1b shows the sending setup in very strong turbulence conditions on the island of La Palma. Vortices and eddies formed by water vapor droplets in the air can be clearly seen in the inset. We were unable to recognize any spatial modes transmitted sent during such atmospheric conditions. The long-time exposure photograph in Fig. 1c was taken under much better atmospheric conditions and shows an $\ell=\pm1$ superposition being transmitted from La Palma to Tenerife, over the edge of the crater wall of the Caldera de Taburiente. The insets show that the double-lobed structure of the mode is clearly visible in the transmitted beam.

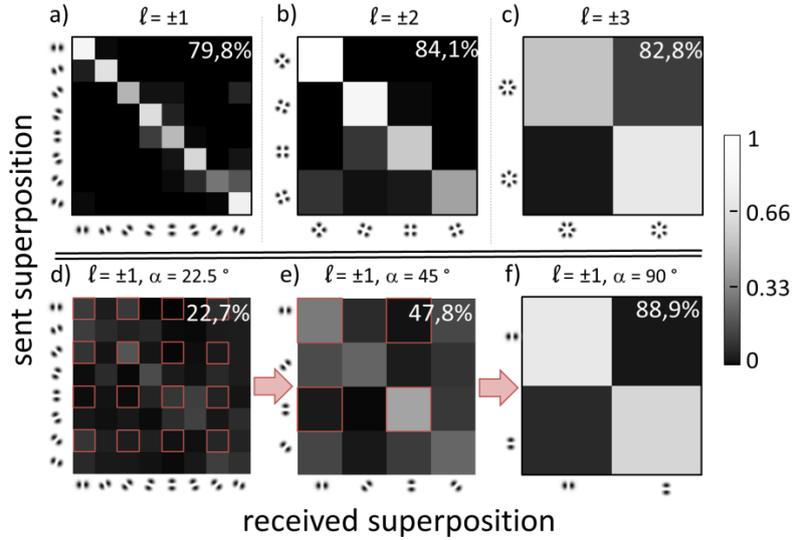

**Figure 3:** Cross talk matrices showing the success probability with which the transmitted OAM mode superpositions were correctly identified at the receiver. **a)-c)** Results for superpositions of $\ell=\pm1$, $\pm2$, and $\pm3$ modes with relative phases of $\Delta\alpha = \frac{\pi}{4}, \frac{\pi}{2},$ and $\pi$ respectively. The different relative phases correspond to different rotations of the superposition structure. The received modes were correctly identified by our detection algorithm with an average success probability of 82%. **d)** During a turbulent night, 8 modes consisting of superpositions of $\ell=\pm1$ were identified with a success probability of only 22.7%. When we restrict ourselves to a subset of these modes with $\Delta\alpha = \frac{\pi}{2}$ or $\pi$ (highlighted with red squares and reanalyzed by the neural network each time), the success probability increases significantly. The ability to resolve all eight of these modes is required for device-independent quantum key distribution (violation of a Bell inequality), four modes are necessary for entanglement-based quantum key distribution (violation of an entanglement witness), and two modes can be used for classical communication with one bit per mode.

To characterize the turbulence, we took long-time exposure photographs of a Gaussian beam shining back from the receiver towards the sender at La Palma and investigated the observed beam spread. Without atmosphere one would expect a Point Spread Function (PSF) with a Full Width Half Maximum (FWHM) of FWHM = λ/D, where D is the effective aperture of the objective. Propagation trough turbulent media enlarges the diameter of the PSF, which leads to a FWHM$_{obs}$, and can be characterized by the Fried parameter r0 that is defined by r0 = λ/FWHM$_{obs}$. We used an Olympus E-M10 camera and an objective with an f-stop of f/5.6 and focal length f = 200mm to record images of the tracking beam at La Palma. The effective aperture of this system was D=35mm and therefore twice as big as the expected r0. During the observation period, the Fried parameter varied between 0.4 cm and 1.3 cm, which is consistent with measurements over the same link from earlier years [40] and is considered to demonstrate strong turbulence.

We performed the measurements over ten successive nights. On four of these nights, we analyzed data for mode superpositions, and on two nights, we recorded pure vortex modes. On the remaining four nights, the weather conditions were too bad in order to recognize any mode structure. This was either because of fog or clouds between the two islands that significantly reduced the received laser intensity, or because of the presence of strong turbulence that significantly deteriorated the mode quality. In these cases, the vertical Fried parameter r0 was below 1 cm. From more details see table 1 in appendix A.

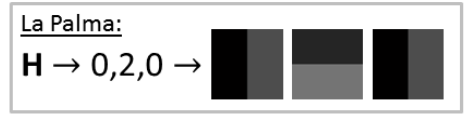

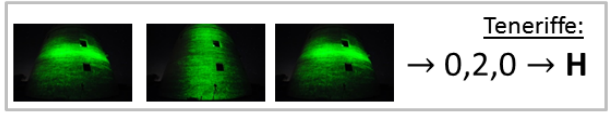

**Figure 4**: Encoding and decoding of a short message with twisted light superpositions. The message "`Hello World!`" is sent letter by letter. Every letter is encoded into three ℓ=±1 superpositions with four different relative phase settings. For example, the letter 'H' is encoded as 0, 2, and 0. Thus every mode corresponds to two bits of information. After 143 kilometers of transmission, the modes are recorded and characterized with an artificial neural network. The same alphabet is used to decode the letter from the mode superpositions. The final recorded message is "`Hello WorldP`". The last letter is a "P" (which is encoded as 1,0,0) instead of a "`!`" (which is encoded as 0,0,0). This error is due to one incorrectly detected mode.

When the Fried parameter r0 was above 1 cm, we were able to discern the OAM modes received at Tenerife. Examples of four of such OAM mode-superpositions as well as four pure vortex modes are shown in Fig. 2. The lobed modal structure is clearly visible for modes with ℓ=±1, ±2, and ±3 in Fig. 2a-d. The relative phase of π introduced in the ℓ=±3 mode superposition is clearly seen to rotate the mode structure by an angle of π/3 in Fig. 2c and 2d. Examples of four different pure vortex modes (ℓ=3, 4, 5, and 7) received after 143km are shown in Figs. 2e-f. As expected, the mode diameter gets larger as the OAM quantum number ℓ is increased, and is seen to approach the size of the telescope wall for ℓ=7.

We used an adaptive pattern recognition algorithm in order to characterize the quality of the received modes. This algorithm was based on an artificial unsupervised neural network, also known as a self-organizing feature map [41, 42]. The idea, which has been used and explained in detail in [33], is to supply the neural network with a training set of images. The images are then characterized automatically according to their respective features. After this training phase, the network can analyze real data in the form of images. The training set consists of images that were sent through the same turbulent link, allowing the algorithm to automatically find a robust characterization of images of modes exposed to turbulence.

Using this detection method, we are able to calculate cross-talk matrices between the sent and received modes. We analyze superposition structures and their angular rotations, which correspond to different relative phases $\Delta\alpha$ between the modes (see Fig. 3, upper row). For ℓ=±1 and eight different settings of $\Delta\alpha = \frac{\pi}{4}$, nearly 80% of the different rotation angles of the structure were correctly identified. For ℓ=±2 with four different settings of $\Delta\alpha = \frac{\pi}{2}$, 84% of the mode images were correctly identified. For superpositions made up of ℓ=±3, two different relative phases $\Delta\alpha = \frac{\pi}{2}$ were identified with nearly 83% certainty. On certain nights, the atmospheric conditions did not allow small phase changes to be identified very well, as is seen by the success rate of 22.7% for ℓ=±1 and $\Delta\alpha = \frac{\pi}{4}$ (Fig. 3d). By restricting ourselves to a smaller subset of these modes, we were able to increase the success rate to 47.8% and 88.9% for $\Delta\alpha = \frac{\pi}{2}$ and $\Delta\alpha = \pi$, respectively. Verifying the presence of OAM entanglement requires one to measure different mode superpositions with at least $\Delta\alpha = \frac{\pi}{2}$. This indicates that in certain conditions where quantum communication would break down, classical communication using OAM states might still be possible. By analyzing intensity images, we estimate the expected visibility in a quantum entanglement experiment to be ~60%, which indicates that the distribution of quantum entanglement over this link encoded in spatial modes is not prevented by atmospheric turbulence (see appendix for more details).

As a final test of the transmission quality, we encoded a short message ('Hello World!') in modes ℓ=±1 with four different relative phases corresponding to $\Delta\alpha = \frac{\pi}{2}$ (see Fig. 4). The message was encoded using a 64-letter alphabet with upper- and lower-case letters, as well as numbers, space and exclamation mark. In this alphabet, every letter needs 6 bits of information for encoding. We encoded each letter by using three consecutively transmitted superposition modes, with each mode in one out of four phase settings. Thus, each mode carries 2 bits of information, which leads to $4^3 = 2^6$ = 64 settings. We recorded five 1-second long exposures of every mode, resulting in a total of 180 images. Out of these, 72 images were chosen at random and used to train the neural network. The network then analyzed the remaining 108 images in order to decode the message. Each letter was redundantly encoded three times as a simple form of error-correction. The network decoded the message as 'Hello WorldP', where the last letter contains one (out of three) wrongly detected mode. The error per letter is 8.33% (1 out of 12 letters are wrong), the error per bit is 1.4% (1 out of 72 bits are wrong). The individual modes were identified correctly with a probability of 76.3%. The complete transmission (including delimiters) took 271 seconds, which corresponds to a speed comparable to that of smoke signaling—the first form of long-distance communication in ancient times, [43] or to that of communication with neutrinos [44].

The quality of the received OAM modes was primarily reduced due to turbulence near the sender—above the Taburiente caldera. The outgoing laser beam was shifted laterally by this turbulence, effectively smearing out the modes received at Tenerife. This first-order effect can be corrected to some extent by using a tip-tilt motor that relies on a beacon laser sent from Tenerife. The correction speed is limited by three primary factors: 1) the physical inter-island distance of 143km, which corresponds to a travel time of 0.5ms, 2) The processing speed of the computer running the correction software, and 3) the tip-tilt motor speed and response time. In order to improve the mode quality in future experiments, one would need to address all of these three points. The physical distance limitation can be addressed by having a second beacon laser closer to the sender. One could illuminate the edge of the caldera wall with a laser beam, thus creating a second "guide star" closer to the sender. This would provide information exclusively about the local turbulence around the sender, which causes a stronger effect on the lateral movement of the received modes than turbulence around the receiver. The second point can be simply addressed by using a fast computer and modern turbulence-compensation software. Finally, a small, two-axis, piezo-controlled mirror placed just before the sending lens could provide fast tip-tilt correction. Taking all these improvements into account, we are believe that the quality of the modes received after 143km of propagation can be improved substantially in future experiments. Furthermore, with this increased pointing stability, performing quantum experiments with the OAM of photons should be feasible [34].

In conclusion, we have investigated the transmission of twisted light modes and their superpositions over a 143-kilometer free-space link between the two Canary Islands, La Palma and Tenerife. Without any active compensation for the effects of turbulence, we find that the relative orientations of the first three higher-order mode superpositions can be distinguished with an average success probability of 82%. This indicates that the long-distance distribution of quantum entanglement of orbital angular momentum modes might be feasible in the future. In order to demonstrate the quality of the link, we transmitted and successfully decoded a short message encoded in the twisted light superpositions. We don't consider this method as real communication but mere the demonstration of the transmission quality of modes. However, the application of state-of-the-art adaptive optics such as those used in simple and efficient intensity-based methods [45] could further improve the link quality, potentially enabling its application in a multiplexing scheme for classical communication [36]. Furthermore, the effective vertical thickness of the atmosphere is 6 kilometers [46], which is well below our link distance, indicating that earth-to-satellite communication with spatially encoded modes is not limited by atmospheric turbulence.

**Acknowlegdements**

The authors thank Z. Sodnik from the European Space Agency (ESA) and J. Carlos from Instituto de Astrofísica de Canarias (IAC) for their support. This work was supported by the European Space Agency (ESA), by the Austrian Academy of Sciences (ÖAW), by the Austrian Federal Ministry of Science, Research and Economy (BMWFW), by the European Research Council (SIQS Grant No. 600645 EU-FP7-ICT) and the Austrian Science Fund (FWF) with SFB F40 (FOQUS).

# Appendix

**Details to Figure 3:**

| mode<br>(date, time) | a) $\ell \pm 1$<br>(12.04.2015, 01:45) | b) $\ell \pm 1$<br>(12.04.2015, 02:45) | c) $\ell \pm 1$<br>(15.04.2015, 23:45) | d) $\ell \pm 2$<br>(16.04.2015, 00:15) | e) $\ell \pm 1$<br>(19.04.2015, 01:45) | f) $\ell \pm 3$<br>(21.04.2015, 02:30) |
|---|---|---|---|---|---|---|
| Δangle | 45° | 22,5° | 22,5° | 22,5° | 45° | 60° |
| exposure time | 0,5 sec | 0,5 sec | 0,5 sec | 1 sec | 0,25 sec | 0,33 sec |
| #(images) | 198 | 1024 | 250 | 133 | 240 | 115 |
| #(training set) | 20 | 120 | 40 | 20 | 20 | 16 |
| duration of series | 5 mins | 23 mins | 21 mins | 8 mins | 9 mins | 3 mins |
| r0 of Zenith | 9,3 cm | 10,2 cm | 15,1 cm | 13,8 cm | 11,6 cm | 11,1 cm |
| r0 of link | - | - | 1,1 cm | 1,1 cm | 1,0 cm | 1,2 cm |
| recognized percentage | 58,1% | 22,7% | 79,8% | 84,1% | 74,1% | 82,8% |

**Table 1**: Cross talk matrices for different OAM superposition settings, with different relative phases. The relative phases correspond to different rotations of the superposition structure. "Δangle" stands for the physical rotations between the different modes, and corresponds to the phase between the superposition. "exposure time" is the time we record the screen with the mode-structure for the data set. "#(images)" is the number of total images of the data set. "#(training set)" is the number of images that were used for training the neural network. "duration of series" is the time between the first and last image of the data set. "r0 of Zenith" is the Fried parameter r0 in the direction of the Zenith, measured at Tenerife by RoboDIMM (Robotic Differential Image Motion Monitor). "r0 of link" is the horizontal Fried parameter between the two islands, over the 143km link, recorded by using a 523nm laser. "recognized percentage" stands for the percentage of correctly identified modes by the neural network in the particular data set.

**Further characterization of mode quality**

We used a set of recorded superposition structures with $\Delta\alpha = \frac{\pi}{2}$, to estimate whether a future quantum experiment is limited by the turbulence. Similar to the method described in [34, 38, 39], we evaluated the intensity distribution depending on the angular position from which we can estimate the achievable visibility of the two mutually unbiased superposition bases and thus if the state is separable or not. We find that even over a measurement time of around 2.5 s per mode an average visibility per basis would be around 60%, thus entanglement would be detectable (if the two-photon coherence is not destroyed in any other way).